\documentclass[12pt]{article}

\usepackage{rotating}
\usepackage{amstext}
\usepackage{amssymb}

\font\twelve=cmbx10 at 15pt
\font\ten=cmbx10 at 12pt

\def\be{\begin{equation}}
\def\ee{\end{equation}}
\def\bq{\begin{eqnarray}}
\def\eq{\end{eqnarray}}
\def\beq{\begin{eqnarray}}
\def\eeq{\end{eqnarray}}
\def\ov{\overline}

\def\epl{e^{(\lambda)}}
\newcommand{\sect}[1]{\setcounter{equation}{0}\section{#1}}

\def\mx{\frac{x^2m^2_\rho}{4}}

\def\Drl{\stackrel{\leftrightarrow}{D}}

\begin{document}

\begin{titlepage}

\begin{center}

{\ten Centre de Physique Th\'eorique\footnote{Unit\'e Propre de
Recherche 7061} - CNRS - Luminy, Case 907}

{\ten F--13288 Marseille Cedex 9 - France }

\vskip 2truecm

{\twelve 
MASS CORRECTIONS OF \\
VECTOR MESON DISTRIBUTION AMPLITUDES}

\bigskip


{\bf 
G. Stoll
}

\vskip 2truecm

{\bf Abstract} 

\medskip

\end{center}

A model-independent way of computing the mass corrections of vector-meson distribution amplitudes is given. The method is based on resummation of (next-to-)leading twist local operator.

\vskip 2truecm



\bigskip

\bigskip


\noindent September 1999


\noindent CPT-99/P.3892

\bigskip

\noindent anonymous ftp : ftp.cpt.univ-mrs.fr

\noindent web : www.cpt.univ-mrs.fr

\pagebreak

\end{titlepage}

\sect{Introduction}

The notion of light-cone wave functions and distribution amplitudes is a well known tool to study the physics of light mesons\cite{BrLe, BrLe2, BrPa}. These concepts were indroduced in the framework of light-cone quantification; they also appear in the parametrization of non-perturbative behavior of light mesons in the infinite momentum frame\cite{CZ}.

The different distributions amplitudes are the basic non-perturbative objects needed for the QCD light-cone sum rules\cite{B}, which is one of the methods to compute exclusive decays of heavy mesons into light mesons (an important way to test the Standard Model and its extensions). In order to obtain more precise results, the distribution amplitudes generate by a non-zero meson mass is usefull.

The most popular model for these distribution amplitudes is the conformal expansion, where they are developped on a (almost) multiplicatively renormalizable basis\cite{BF, BBK}. In this framework, the meson mass corrections are obtained via the exact equation of motion\cite{BBS}. But other models exist (renormalon model in \cite{An}, instanton model in \cite{Wei}), therefore a model-independant way to compute the meson mass corrections is needed.

The method used in this letter consist of expressing the distribution amplitudes generated by the mass of a vector meson in term of the leading twist distribution amplitude, keeping only the (next-to-)leading terms in the Operators Products Expansion on the light-cone\cite{BP}. This idea was introduced in \cite{BBS} in the case of a scalar theory.

The paper is organized as follows. After the introduction of the usefull notations, the different notions of twist is explained, on which the method of computing mass correction is based.

Then the case of chiral even and chiral odd two-points distribution are treated.

%
%

\sect{Notations}

The two-points distrubtion amplitudes considered are extracted from this kind of matrix element~:
\be
\left\langle 0 |\ov{u}(x)[x,-x] \Gamma d(-x)|\rho \right\rangle \label{2matel}
\ee

Where $x$ is almost on the light cone, $v \in [0,1]$, $\Gamma$ any kind of product of $\gamma_{\mu}$ matrices and $[x,y]$ a path-ordered gauge factor along the straight line connecting $x$ and $y$ :
\be
[x,y]={\rm P}\exp \left[ ig\int^1_0 dt (x-y)_\mu A^\mu (tx+(1-t)y)\right]
\ee

This gauge factor is a way to introduce interaction (see \cite{Bali}), it ensure the gauge invariance of these non-local matrix elements (I will omit to write it sometimes).

These matrix elements depend on three vectors : 

\noindent - $P_{\mu}$ : momentum of the $\rho$-meson

\noindent - $\epl_{\mu}$ : polarization vector of the $\rho$-meson

\noindent - $x_{\mu}$

With the relations :
\bq 
P^2&=&m^2_{\rho} \\ \epl\cdot\epl&=&-1 \\ P\cdot\epl&=&0
\eq

%
%

\sect{Definitions of twist \label{deftw}}

The different contributions to the matrix elements \ref{2matel} are classified by their "twist", but there are two different
definitions of this concept: 

For local operators, the "twist" means dimension minus Lorentz-spin. To extract the contribution of a definite twist for non-local matrix elements like \ref{2matel}, an expansion in local operator (OPE) has to be done and all local operators which have a definite twist in the sense given above must be re-summed. This will give
the non-local term which has a definite power in $x^2$. To understand how this
definition of "twist" means "power in $x^2$", see \cite{BP}. From now on, this definition will be called "theoretical twist".

The second definition is built in analogy to partons distributions. In the latter case, the different structures of non-local matrix elements are separated with their different powers of $Q$ (the hard momentum transfer) in deep inelastic scattering : a term of
twist $t$ contributes to the inclusive cross section with coefficients which contain $t-2$ or more power of $1/Q$. A good description of this classification can be found in \cite{JJ}. From now on, this definition will be called "physical twist".

In this letter, the distributions amplitudes which are generated by meson mass corrections (these are non-leading "physical" twist distribution amplitudes) are computed in the leading (and next-to-leading) "theoretical" twist expansion. 

%
%

\sect{Meson mass correction}

%

\subsection{chiral-even case}

The mass of the $\rho$-meson induces some non-leading (physical-)twist structure
for the non-local matrix element:
\begin{eqnarray}
\lefteqn{\langle 0|\bar u(x) \gamma_{\mu} [x,-x] 
d(-x)|\rho^-(P,\lambda)\rangle = }\makebox[2cm]{\ } \nonumber \\
&=& f_{\rho} m_{\rho} \left[ P_{\mu}
\frac{e^{(\lambda)}\cdot x}{P \cdot x}
\int_{0}^{1} \!du\, e^{i \xi P \cdot x} \left(A^{(e)}(u)+
\frac{x^2 m_\rho^2}{4}\hat{A}^{(e)}(u)\right) \right. 
\nonumber \\
& &{}+ \left(e^{(\lambda)}_{\mu}-P_\mu\frac{\epl\cdot x}{P\cdot x}\right)
\int_{0}^{1} \!du\, e^{i \xi P \cdot x} \left(B^{(e)}(u) 
+\frac{x^2 m_\rho^2}{4}\hat{B}^{(e)}(u)\right)\nonumber \\
& & \left.{}- \frac{1}{2}x_{\mu}
\frac{e^{(\lambda)}\cdot x }{(P \cdot x)^{2}} m_{\rho}^{2}
\int_{0}^{1} \!du\, e^{i \xi P \cdot x} C(u)^{(e)} \right] \label{tpmasscor}
\end{eqnarray}
with $\xi=2u-1$.

Some wave functions  would
disappear if the mass of the $\rho$ were $0$. Hence mass corrections are higher
(physical-)twist contributions to non-local matrix elements ($\hat{A}^{(e)}(u)$ and $C^{(e)}(u)$ are twist 4, $\hat{B}^{(e)}$ is twist 5).

The idea is to compute the mass corrections in the leading theoretical twist approximation. In this part, the case with the two-point chiral even distributions is computed.
The "brute force" method, i. e. the expansion of the distributions in moments ($x\rightarrow 0$) can be used. It gives :
\bq
\lefteqn{\langle 0|\ov{u}\gamma_\mu(i\Drl x)^n d|\rho^-(P,\lambda)\rangle
=} \nonumber \\
&=&P_\mu\frac{\epl\cdot x}{P\cdot x}\left[(Px)^nM_n^A-\frac{x^2m_\rho^2}{4}
n(n-1)\hat{M}^A_{n-2}(px)^{n-2}\right] \nonumber \\
& &{}+ \left(e^{(\lambda)}_{\mu}-P_\mu\frac{\epl\cdot x}{P\cdot x}\right)
\left[(Px)^nM_n^B-\frac{x^2m^2_\rho}{4}n(n-1)\hat{M}^B_{n-2}(Px)^{n-2}\right]
\nonumber \\
& &{}-\frac{1}{2}x_\mu\frac{\epl\cdot x}{(P\cdot x)^2}m_\rho^2(Px)^nM_n^c
\eq
 where the moments $M_n^A$, $M_n^B$, $M_n^C$ $\hat{M}_n^A$ and $\hat{M}_n^B$ are defined like in the following way: $M_n=\int_0^1du\xi^nf(u)$.

The leading (theoretical-)twist of the operators in the left hand side can be written. The condition of symmetry and zero trace gives :
\bq
\lefteqn{\langle 0|\ov{u}\gamma_\mu(i\Drl x)^n d|\rho^-(P,\lambda)\rangle=} \nonumber \\
&=&\left\{{} P_\mu\frac{\epl\cdot x}{P\cdot x}(Px)^n+\frac{1}{n+1}
\left(e^{(\lambda)}_{\mu}-P_\mu\frac{\epl\cdot x}{P\cdot x}\right)
\left[(Px)^n-\mx\frac{n(n-1)}{n+1}(Px)^{n-2}\right]\right. \nonumber \\
& &{}-\left. \mx P_\mu\frac{\epl\cdot x}{P\cdot x}(Px)^{n-2}
n\frac{(n-1)^2}{(n+1)^2}-\frac{1}{2}x_\mu\frac{\epl\cdot x}{(P\cdot x)^2}
(Px)^nm_\rho^2\frac{(n-1)n}{(n+1)^2}\right\}\langle\langle O_n\rangle\rangle
\nonumber \label{defOn} \\
\eq

The comparison of these two equations implies :
\bq
M_n^A&=&\langle\langle O_n\rangle\rangle \label{ltmom1} \\
M_n^B&=&\frac{1}{n+1}\langle\langle O_n\rangle\rangle \\
\hat{M}^A_{n-2}&=&\frac{n-1}{(n+1)^2}\langle\langle O_n\rangle\rangle \\
\hat{M}^B_{n-2}&=&\frac{1}{(n+1)^2} \langle\langle O_n\rangle\rangle \\
M_n^C&=&\frac{(n-1)n}{(n+1)^2}\langle\langle O_n\rangle\rangle \label{ltmom5}
\eq

There is another method to get the leading (theoretical-)twist part of the
distributions, giving directly integrated equations
between the different wave functions. The different relations for
non-local operators in appendix \ref{app:oprel} are used: they give integrated forms for the leading
twist part of two-point operators.

One takes equation \ref{BBformula}:

\be
\left[\ov{\psi}(x)\gamma_\alpha\psi(-x)\right]_{\text{sym.}}=\int_0^1 du
\frac{\partial}{\partial x_\alpha}\ov{\psi}(ux)\hat{x}\psi(-ux) \label{vopsym}
\ee
where $\hat{x}=x^\nu\gamma_\nu$. The subscript "sym." means that in the Operator Product Expansion (local expansion around $x=0$), only the operators which have
symmetrical Lorentz-indices are taken. But for the leading theoretical twist, the traces has also to be removed, so the leading twist part of the right hand side of \ref{vopsym} is also taken :
\be
\left[\ov{\psi}(x)\gamma_\alpha\psi(-x)\right]_{\text{l. t.}}=\int_0^1 du
\frac{\partial}{\partial x_\alpha}\left[\ov{\psi}(ux)\hat{x}\psi(-ux) \right]_{\text{l. t.}}\label{voplt}
\ee

The condition for the leading twist part of $\ov{\psi}(ux)\hat{x}\psi(-ux)$ is the following (equation \ref{ltxhatapp}) :
\be
\frac{\partial}{\partial x_\alpha\partial x^\alpha}\left[\ov{\psi}(x)\hat{x}
\psi(-x)\right]_{\text{l. t.}}=0 \label{ltxhat}
\ee
 
Now the matrix element of $\ov{\psi}(ux)\hat{x}\psi(-ux)$ can be taken. The parameterization of the leading twist part up to $m_\rho^2$ corrections is the following:

\bq
\lefteqn{\langle 0|\left[\bar u(x) \hat{x} [x,-x] 
d(-x)\right]_{\text{l. t.}}|\rho^-(P,\lambda)\rangle =} \nonumber \\
&=&(\epl\cdot x)\int_0^1du g(u)\left[e^{i\xi Px}+\mx f_\xi(Px)+O(m_\rho^4)\right]
\label{xhatpar}
\eq
where $g(u)$ is the leading twist wave functions in the limit $m_\rho\rightarrow
 0$.

The condition \ref{ltxhat} can be applied; it gives (after some calculation) :
\be
\int_0^1 du g(u)\left[3f_\xi(Px)+(Px)f'_\xi(Px)-\xi^2e^{i\xi Px}\right]=0
\label{condfxi}
\ee

The following ansatz for $f_\xi(Px)$ can be done (having in mind the case of scalar field\cite{BBS}):
\be
f_\xi=a(\xi)\int_0^1 dv b(v)e^{i\xi vPx}
\ee

Then equation \ref{condfxi} gives
\bq
a(\xi)&=&\xi^2 \\
b(v) &=&v^2 
\eq

This can be written in this form :
\bq
\lefteqn{\langle 0|\left[\bar u(x) \hat{x} [x,-x] 
d(-x)\right]_{\text{l. t.}}|\rho^-(P,\lambda)\rangle =} \nonumber \\
&=&\int_0^1du g(u)\left[(\epl\cdot x)e^{i\xi Px}\right]_{\text{l. t.}}
\eq
with
\bq
\lefteqn{\left[(\epl\cdot x)e^{i\xi Px}\right]_{\text{l. t.}}=} \nonumber \\
&=&(\epl\cdot x)\left[e^{i\xi Px}+\mx\xi^2\int_0^1 dv v^2 e^{iv\xi Px}\right]
\eq

The equation \ref{voplt} can be applied. After some algebra, one gets the distributions $A^{(e)}(u)$, $B^{(e)}(u)$, $C^{(e)}(u)$, $\hat{A}^{(e)}(u)$ and
$\hat{B}^{(e)}(u)$ defined in \ref{tpmasscor} in the leading (theoretical-) twist approximation :
\bq
A^{(e)}(u)&=&g(u) \label{Aelt}\\
\left.\int_0^1duB^{(e)}(u)e^{i\xi Px}\right)_{\text{l. t.}}&=&\int_0^1du g(u)\int_0^1dte^{it\xi Px} \\
\left.\int_0^1duC^{(e)}(u)e^{i\xi Px}\right)_{\text{l. t.}}&=&\int_0^1dug(u)\left[e^{i\xi Px}-3\int_0^1dt
e^{it\xi Px}+2\int_0^1dt\int_0^1dve^{ivt\xi Px}\right] \nonumber \\ \\
\left.\int_0^1du\hat{A}^{(e)}(u)e^{i\xi Px}\right)_{\text{l. t.}}&=&
\int_0^1dug(u)\xi^2\int_0^1dt\,t^2\left[e^{it\xi Px}-2\int_0^1dvv^2e^{ivt\xi Px}
\right] \\
\left.\int_0^1du\hat{B}^{(e)}(u)e^{i\xi Px}\right)_{\text{l. t.}}&=&
\int_0^1dug(u)\int_0^1dt\xi^2\int_0^1v^2dve^{ivt\xi Px} \label{Bhatelt}
\eq

There is no index "l. t." for $A^{(e)}(u)$ because this distribution contains only the leading twist part of the matrix element $\langle 0|\bar u(x) \gamma_\mu [x,-x] 
d(-x)|\rho^-(P,\lambda)\rangle$. So these last equations show that all the distributions can be computed when one has $A^{(e)}(u)$, in the leading (theoretical-)twist approximation. They reproduced exactly the relations between moments (equations \ref{ltmom1} to \ref{ltmom5}). The advantage is that they are directly in an integrated form.

\subsubsection{Adding twist 3 and twist 4 terms}

The idea is to add contributions of (theoretical-)twist 3 and 4 without the gluonic contribution. In the appendix \ref{app:oprel}, there is the following operator relation (\ref{dertot}) :
\bq
\ov \psi(x)\hat{x}\psi(-x)&=&\left[\ov \psi(x)\hat{x}\psi(-x)\right]_
{\text{l. t.}}-\frac{1}{4}x^2\int_0^1dv v^2\partial^2
\ov \psi(vx)\hat{x}\psi(-vx) \nonumber \\
& &{}+ \text{operators with gluons}+\ldots \label{txtdertot}
\eq

The matrix element of the term with a total derivative, gives
\bq
\lefteqn{
-\frac{1}{4}x^2\int_0^1dv v^2\langle 0|\partial^2\ov u(vx)\hat{x}[vx,-vx]d(-vx)|
\rho(P,\lambda)\rangle =} \nonumber \\
&=&m^2_\rho\frac{1}{4}x^2\int_0^1dv v^2\langle 0|\ov u(vx)\hat{x}[vx,-vx]d(-vx)|
\rho(P,\lambda)\rangle
\eq

The $O(m_\rho^0)$ term (equation \ref{xhatpar}) in the parameterization of the matrix element in the right hand side implies :
\bq
\lefteqn{
-\frac{1}{4}x^2\int_0^1dv v^2\langle 0|\partial^2\ov u(vx)\hat{x}[vx,-vx]d(-vx)|
\rho(P,\lambda)\rangle =} \nonumber \\
&=&(\epl\cdot x)\mx\int_0^1 du g(u)\int_0^1dv v^2e^{iv\xi Px}
\eq

So the leading twist part with the total derivative term is :
\bq
\lefteqn{\langle 0|\left[\bar u(x) \hat{x} [x,-x] 
d(-x)\right]_{\text{l. t. + tot. der.}}|\rho^-(P,\lambda)\rangle =} \nonumber \\
&=&\int_0^1du g(u)\left[(\epl\cdot x)e^{i\xi Px}\right]_
{\text{l. t. + tot. der.}}
\eq
with
\bq
\lefteqn{\left[(\epl\cdot x)e^{i\xi Px}\right]_{\text{l. t. + tot. der.}}=} \nonumber \\
&=&(\epl\cdot x)\left[e^{i\xi Px}+\mx(\xi^2+1)\int_0^1 dv v^2 e^{iv\xi Px}\right]
\label{totdercor}
\eq

This last form can be used to give some twist 4 corrections to equation \ref{Aelt}
to \ref{Bhatelt}.

Now the twist 3 part without gluon can be added. The matrix element of the equations \ref{twist3} is taken, suppressing the gluonic terms :
\bq
\lefteqn{\langle 0|\left[\bar u(x) \gamma_\mu [x,-x] 
d(-x)\right]_{\text{t3}}|\rho^-(P,\lambda)\rangle =} \nonumber \\ 
&=&-i\epsilon_{\mu\nu\alpha\beta}\int_0^1dt\,t x^\nu(-iP^\alpha)\langle 0|\bar u(tx) \gamma_\beta\gamma_5 [tx,-tx] 
d(-tx)|\rho^-(P,\lambda)\rangle \nonumber \\ \\
\lefteqn{\langle 0|\left[\bar u(x) \gamma_\mu \gamma_5[x,-x] 
d(-x)\right]_{\text{t3}}|\rho^-(P,\lambda)\rangle =} \nonumber \\ 
&=&-i\epsilon_{\mu\nu\alpha\beta}\int_0^1dt\,t x^\nu(-iP^\alpha)\langle 0|\bar u(tx) \gamma_\beta [tx,-tx] 
d(-tx)|\rho^-(P,\lambda)\rangle \nonumber \\
\eq

Combining these two equations, replacing the right hand side of the second one with the
parameterization of \ref{tpmasscor}, one gets:
\bq
\lefteqn{\langle 0|\left[\bar u(x) \gamma_\mu [x,-x] 
d(-x)\right]_{\text{t3}}|\rho^-(P,\lambda)\rangle =} \nonumber \\ 
&=&(\epl_\mu(P\cdot x)^2+P_\mu(\epl\cdot x)(P\cdot x)\int_0^1du \int_0^1 dt\,t
\int_0^1 dv v^2 B^{(e)}(u)e^{ivt\xi Px} \nonumber \\
& &{}+m^2_\rho\int_0^1du\int_0^1 dt t\int_0^1 dv v^2 e^{ivt\xi Px}\times
\nonumber \\
& &{}\;\times \left[(\epl_\mu(P\cdot x)^2+P_\mu(\epl\cdot x)(P\cdot x)
\frac{t^2x^2v^2}{4}\hat{B}^{(e)}(u)+\left(x^2\epl_\mu-x_\mu(\epl\cdot x)\right)
B^{(e)}(u)\right] \nonumber \\
\eq

Including the leading twist term, the total derivative term (\ref{totdercor}) and this last equation, equations for $A^{(e)}(u)$, $B^{(e)}(u)$, $C^{(e)}(u)$, $\hat{A}^{(e)}(u)$ and
$\hat{B}^{(e)}(u)$ can be obtained:
\bq
A^{(e)}(u)&=&g(u) \label{Aet234}\\
\int_0^1duB^{(e)}(u)e^{i\xi Px}&=&\int_0^1du g(u)\int_0^1dte^{it\xi Px}\nonumber
\\
& &{}-(P\cdot x)^2\int_0^1du\int_0^1dt\,t\int_0^1dv\,v^2e^{ivt\xi Px}B^{(e)}(u)
\nonumber \\ \\
-\frac{1}{2}\frac{1}{(P\cdot x)^2}\int_0^1duC^{(e)}(u)e^{i\xi Px}&=&\int_0^1du\frac{\xi^2+1}{\xi}g(u)\int_0^1dt\,t^2\int_0^1dvv^2e^{ivt\xi Px} \nonumber \\ 
& &{}-\int_0^1du B^{(e)}(u)\int_0^1dt\,t\int_0^1dv\,v^2e^{ivt\xi Px} \nonumber \\
\\
\int_0^1du\hat{B}^{(e)}(u)e^{i\xi Px}&=&
\int_0^1dug(u)(\xi^2+1)\int_0^1dt\,t^2\int_0^1v^2dve^{ivt\xi Px} \nonumber \\
& &{}+4\int_0^1duB^{(e)}(u)\int_0^1dt\,t\int_0^1dv\,v^2e^{ivt\xi Px} \nonumber
\\
& &{}-(P\cdot x)^2\int_0^1du\hat{B}^{(e)}(u)\int_0^1dt\,t^3
\int_0^1dv\,v^3e^{itv\xi Px} \nonumber \\ \label{Bhatet234}
\eq
\bq\lefteqn{\int_0^1du\hat{A}^{(e)}(u)e^{i\xi Px}=} \nonumber \\
& &{}
\int_0^1dug(u)(\xi^2+1)\int_0^1dt\,t^2\int_0^1dv\,v^2\left(i\xi(P\cdot x)+1\right)
t^2e^{ivt\xi Px} \nonumber \\
& &{}+4\int_0^1duB^{(e)}(u)\int_0^1dt\,t\int_0^1dv\,v^2e^{ivt\xi Px} 
\nonumber \\
\eq

Theoretically, all the chiral-even distributions $B^{(e)}(u)$, $C^{(e)}(u)$, $\hat{A}^{(e)}(u)$ can be computed once the leading (physical-)twist contribution $A^{(e)}(u)$ is known. But, practically, it quite long and hard if one has a conformal expansion for $A^{(e)}(u)$ (except if only the asymptotic part of the wave function is taken, see \cite{BBS}). The advantage of these equations is that they are independent of the model used to computed $A^{(e)}(u)$ (another expansion can be used as the conformal one). They can be the basis of a numerical computation if one has discrete values for $A^{(e)}(u)$.

%

\subsection{Chiral odd distributions}
For the different wave functions
\begin{eqnarray}
\lefteqn{\hspace*{-1.5cm}\langle 0|\bar u(x) \sigma_{\mu \nu} [x,-x] 
d(-x)|\rho^-(P,\lambda)\rangle =} \nonumber \\
&=& i f_{\rho}^{T} \left[ ( e^{(\lambda)}_{\mu}P_\nu -
e^{(\lambda)}_{\nu}P_\mu )
\int_{0}^{1} \!du\, e^{i \xi P \cdot x} \left(A^{(o)}(u) 
+\frac{m^2_\rho x^2}{4}\hat{A}^{(o)}(u)\right)\right. 
\nonumber \\
& &{}+ (P_\mu x_\nu - P_\nu x_\mu )
\frac{e^{(\lambda)} \cdot x}{(P \cdot x)^{2}}m_{\rho}^{2} 
\int_{0}^{1} \!du\, e^{i \xi P \cdot x}\left(B^{(o)}(u) 
+\frac{m^2_\rho x^2}{4}\hat{B}^{(o)}(u)\right)  
\nonumber \\
& & \left.{}+ \frac{1}{2}
(e^{(\lambda)}_{\mu} x_\nu -e^{(\lambda)}_{\nu} x_\mu) 
\frac{m_{\rho}^{2}}{P \cdot x} 
\int_{0}^{1} \!du\, e^{i \xi P \cdot x} C^{(o)}(u) \right],
\end{eqnarray}
the formulaes of the appendix can be used to obtains integrated equations in terms of the leading-twist distribution $A^{(o)}$:
\bq
\lefteqn{\int_0^1due^{i\xi Px}\hat{A}^{(o)}(u)=} \nonumber \\
&=&\int_u^1duE(u)\int_0^1dt\int_0^1dv\,v^3e^{ivt\xi Px}
\left[-\frac{2}{tv^3(Px)^3}
-\frac{2}{v^3(Px)^3}+2\frac{i\xi}{v^2(Px)^2}\right. \nonumber \\
& &{}\left.+2\frac{i\xi t}{v^2(Px)^2}
-3\frac{(i\xi)^2t}{v(Px)}+2\frac{i\xi)^2t^2}{v(Px)}-(i\xi)^3t^2+\frac{t}{v(Px)}
+t^2(i\xi)\right] \nonumber \\
\eq
\bq
\lefteqn{\int_0^1due^{i\xi Px}C^{(o)}(u)=} \nonumber \\
&=&\frac{1}{2}(P\cdot x)\int_0^1duE(u)\int_0^1dt\int_0^1dv\,v^2e^{itv\xi Px}
\left[\frac{4}{tv^2(Px)}-4\frac{i\xi}{v(Px)}-2(i\xi)^2t+2t\right]\nonumber \\
\eq
\bq
\lefteqn{\int_0^1due^{i\xi Px}B^{(o)}(u)=} \nonumber \\
&=&(P\cdot x )\int_0^1duE(u)\int_0^1dt\int_0^1dv\,v^2e^{itv\xi Px}
\left[-\frac{1}{tv^2(Px)}+\frac{1}{v^2(Px)}+\frac{i\xi}{v(Px)}\right.\nonumber \\
& &{}-\left.
\frac{i\xi t}{v(Px)}+\frac{(i\xi)^2t}{2}-(i\xi)^2t^2-\frac{t}{2}\right]
\eq
where
\be
E(u)\equiv\frac{1}{2}i\frac{d}{du}A^{(o)}(u)
\ee
$\hat{B}^{(e)}(u)$ has not been computed because it is already a $O(m^4_\rho)$ contribution.

%

\section{Conclusion}

This letter describes a model-independant way to extract the mass corrections of vector meson distribution amplitude from the leading twist ones. Although the only the (next-to-)leading twist {\it local} operators are taken into account, the method can be extended. In that case, one would obtain integrated 
equations for meson mass corrections in function of leading and non-leading twist (two and three points) distribution amplitude.

The integrated equation of this letter are quite hard to solve if conformal expansion model for the leading twist distribution amplitude is taken (like in \cite{BBS}); but these equations can be the basis for numerical estimates, especially when an alternative model for the leading twist distribution amplitude is used (or if the distribution amplitude is directely taken form experiemntal results).

\section*{Acknowledgments}

This work is supported by Schweizerischer Nationalfond. I am grateful to V. M. Braun and P. Ball for introducing me to the subject.


\begin{appendix}


\sect{Operators relations for different theoretical twist \label{app:oprel}}

This appendix contains some relations taken from \cite{BaBr}, \cite{BB96} and \cite{BBK}, which help to isolate the different twist parts of two-points non-local operators.

\subsection{Chiral-even operator}

Consider the non-local operators $\ov{\psi}(x)\gamma_\alpha\psi(-x)$. The local expansion around $x=0$ can be written in the following way :
\be
\ov{\psi}(x)\gamma_\alpha\psi(-x)=\sum_{n=0}^\infty x_{\mu_1}\ldots x_{\mu_n}
\frac{1}{n!}\ov{\psi}\Drl_{\mu_1}\ldots\Drl_{\mu_n}\gamma_\alpha\psi
\ee

In order to get the leading twist contribution, the symmetrization has to be done over all indices and subtract the traces :
\be
\left[\ov{\psi}(x)\gamma_\alpha\psi(-x)\right]_{\text{l. t.}}\equiv\sum_{n=0}^\infty x_{\mu_1}\ldots x_{\mu_n}
\frac{1}{n!}\ov{\psi}\left(\Drl_{\mu_1}\ldots\Drl_{\mu_n}\gamma_\alpha\right)
_{\text{sym}}\psi-\text{traces}
\ee

The symmetrization has an integrated solution :
\begin{eqnarray}
\lefteqn{\Big[\bar \psi(x)\gamma_\alpha \psi(-x)\Big]_{\text{sym}} \equiv}
\nonumber\\
&\equiv &
\sum_{n=0}^\infty \frac{x^{\mu_1}\ldots x^{\mu_n}}{n!}
\bar \psi(0)\Bigg\{\frac{1}{n+1}\Drl_{\mu_1}\ldots \Drl_{\mu_n}\gamma_\alpha 
+\frac{n}{n+1}\Drl_{\alpha}\Drl_{\mu_1}\ldots \Drl_{\mu_{n-1}}\gamma_{\mu_n}
\Bigg\} \psi(0)\,.\makebox[1cm]{\ } \nonumber \\
\label{defsym} \\
 & =&
\int_0^1 dv\,\frac{\partial}{\partial x_\alpha}\bar \psi(vx)\hat{x} 
\psi(-v x)
\label{BBformula}
\eq
with $\hat{x}\equiv\gamma_\mu x^\mu$.

In order to have really the leading twist part of  $\ov{\psi}(x)\gamma_\alpha\psi(-x)$, the leading twist part of $\bar \psi(x)\hat{x} \psi(-x)$ is needed. If one does a local expansion of this latter non-local operator, one will already have symetrized local operators. But traces has alos to be removed; this can be expressed as a differential equation :
\be
\frac{\partial}{\partial x_\alpha\partial x^\alpha}\left[\ov{\psi}(x)\hat{x}
\psi(-x)\right]_{\text{l. t.}}=0 \label{ltxhatapp}
\ee
which has the formal solution :
\bq
\left[\ov{\psi}(x)\hat{x}\psi(-x)\right]_{\text{l. t.}}&=&\ov{\psi}(x)\hat{x}\psi(-x)
\nonumber \\
& &{}+\sum_{n=1}^\infty\int_0^1\left(-\frac{1}{4}x^2\right)^n
\frac{\left(\frac{1-t}{t}\right)^{n-1}}{(n-1)!n!}
\left(\frac{\partial^2}{\partial x^\alpha\partial x_\alpha}\right)^n
\ov{\psi}(tx)\hat{x}\psi(-tx) \nonumber \\ \label{solltxhat}
\eq

Equations for the twist 3 part of $\ov \psi(x)\gamma_\alpha\psi(-x)$  are the following :

\begin{eqnarray}
 \Big[\bar \psi(x)\gamma_\mu \psi(-x)\Big]_{\text{twist 3}} &=&
{}-g_s\int_0^1 \!du\int_{-u}^u \!dv\,\bar \psi(ux)
\left[
u\tilde G_{\mu\nu}(vx)x^\nu \hat{x}\gamma_5\right. \nonumber \\ 
& &{} -ivG_{\mu\nu}(vx)x^\nu \hat{x}
\Big]\psi(-ux)
\nonumber\\
&&{}+i\epsilon_{\mu}^{\phantom{\mu}\nu\alpha\beta}\int_0^1 udu\, 
x_\nu\partial_\alpha
\Big[\bar \psi(ux)\gamma_\beta\gamma_5 \psi(-ux)\Big]\,,
\nonumber\\
\Big[\bar \psi(x)\gamma_\mu\gamma_5 \psi(-x)\Big]_{\text{twist 3}} &=&
-g_s\int_0^1 \!du\int_{-u}^u \!dv\,\bar \psi(ux)
\left[
u\tilde G_{\mu\nu}(vx)x^\nu\hat{x}\right. \nonumber \\ & &{}-ivG_{\mu\nu}(vx)x^\nu\hat{x}\gamma_5
\Big]\psi(-ux)
\nonumber\\
&&{}+i\epsilon_{\mu}^{\phantom{\mu}\nu\alpha\beta}\int_0^1 udu\, 
x_\nu\partial_\alpha
\Big[\bar \psi(ux)\gamma_\beta \psi(-ux)\Big]\,,
\label{twist3}
\end{eqnarray}
where $G_{\mu\nu}$ is the gluon field strength, $\tilde G_{\mu\nu}
=(1/2)\epsilon_{\mu\nu\alpha\beta}G^{\alpha\beta}$, and
$\partial_\alpha$ is the derivative over the total translation :
\begin{equation}
\partial_\alpha\Big[\bar \psi(ux)\gamma_\beta \psi(-ux)\Big] \equiv 
\left.\frac{\partial}{\partial y_\alpha}
\Big[\bar \psi(ux+y)\gamma_\beta \psi(-ux+y)\Big]\right|_{y\to 0}. \label{totder} 
\end{equation}

These equations come from \cite{BB96} and can be obtained with the equation of motion, neglecting quark masses.

Another useful formula is the equation (5.13) in \cite{BaBr} which can be written like this :
\be
\frac{\partial}{\partial x_\alpha\partial x^\alpha}\ov \psi(x)\hat{x}\psi(-x)=
-\partial^2\ov \psi(x)\hat{x}\psi(-x) + \text{operators with gluons}
\ee

The expansion of \ref{solltxhat} at the order $x^2$, gives
\bq
\ov \psi(x)\hat{x}\psi(-x)&=&\left[\ov \psi(x)\hat{x}\psi(-x)\right]_
{\text{l. t.}}+\frac{1}{4}x^2\int_0^1dv\frac{\partial^2}{\partial x_\alpha
\partial x^\alpha}\ov \psi(vx)\hat{x}\psi(-vx) + \ldots \nonumber \\ 
&=&\left[\ov \psi(x)\hat{x}\psi(-x)\right]_
{\text{l. t.}}-\frac{1}{4}x^2\int_0^1dv v^2\partial^2
\ov \psi(vx)\hat{x}\psi(-vx) \nonumber \\
& &{}+ \text{operators with gluons}+\text{twist 3}+\ldots \label{dertot}
\eq

This last equation gives a twist 4 contribution with a total derivative (defined in \ref{totder}).

%

\subsection{Chiral-odd operator} 

Consider the non-local operators $\ov{\psi}(x)\sigma_{\alpha\beta}\psi(-x)$. The following equation can be used\cite{Bun}:
\bq
\bar \psi(x)\sigma_{\alpha\beta} \psi(-x)=
\int_0^1 u dv\,\frac{\partial}{\partial x_\beta}\bar \psi(vx)
\sigma_{\alpha\nu}x^\nu 
\psi(-v x)-(\alpha \leftrightarrow \beta)+O(\text{twist }3) \nonumber \\
\eq
to extract the most symetrical part of $\ov{\psi}(x)\sigma_{\alpha\beta}\psi(-x)$:
\begin{eqnarray}
\lefteqn{\Big[\bar \psi(x)\sigma_{\alpha\beta} \psi(-x)\Big]_{\text{sym}} =} \nonumber
\\
 & =&
\int_0^1 v dv\,\frac{\partial}{\partial x_\beta}\bar \psi(vx)
\sigma_{\alpha\nu}x^\nu 
\psi(-v x)-(\alpha \leftrightarrow \beta)
\label{BBformulasig}
\eq
(it reproduces the result in \cite{Laz}).

In \cite{BBK}, the equations for the leading twist part of $\bar \psi(x)\sigma_{\alpha\nu}x^\nu \psi(-x)$ can be founds~:
\bq
\frac{\partial}{\partial x_\alpha}\left[\bar \psi(x)\sigma_{\alpha\nu}x^\nu \psi(-x)\right]_{\text{l. t.}}&=&0 \\
\frac{\partial^2}{\partial x_\rho \partial x^\rho}\left[\bar \psi(x)\sigma_{\alpha\nu}x^\nu \psi(-x)\right]_{\text{l. t.}}&=&0
\eq

Like in equation \ref{dertot}, the contribution of total derivative can be added :
\bq
\bar \psi(x)\sigma_{\alpha\nu}x^\nu \psi(-x)&=&
\left[\bar \psi(x)\sigma_{\alpha\nu}x^\nu \psi(-x)\right]_{\text{l. t.}} -\frac{1}{4} x^2x^\nu\int_0^1 dt\,t^2 \partial^2\left(\bar \psi(tx)\sigma_{\alpha\nu}\psi(-tx)\right) \nonumber \\
& &{}+\text{operators with gluons}+\ldots 
\eq


\end{appendix}

\end{document}